\documentclass[showpacs,10pt,aps,prd,reprint,onecolumn,notitlepage,nofootinbib,superscriptaddress]{revtex4-1}

\usepackage{hyperref}
\usepackage{amsmath}
\usepackage{mathtools}
\usepackage{bm}
\usepackage{amssymb}
\usepackage{graphicx}
\usepackage[]{subfigure}
\usepackage[usenames,dvipsnames,svgnames,table]{xcolor}
\usepackage{filecontents}

\hypersetup{
    bookmarks=true,         % show bookmarks bar?
    pdftoolbar=true,        % show Acrobat's toolbar?
    pdfmenubar=true,        % show Acrobat's menu?
    pdffitwindow=true,     %window fit to page when opened
    pdfstartview={FitH},     %fits the width of the page to the window
    pdftitle={My title},     %title
    pdfauthor={author},      %author
    pdfsubject={Subject},    %subject of the document
    pdfcreator={Creator},    %creator of the document
    pdfproducer={Producer},  %producer of the document
    pdfkeywords={keyword1} %{key2} {key3}, % list of keywords
    pdfnewwindow=true,       %links in new window
    colorlinks=true ,        %false: boxed links; true: colored links
    linkcolor=blue  ,        %color of internal links (change box color with linkbordercolor)
    citecolor=blue,      %color of links to bibliography
    filecolor=green,       %color of file links
    urlcolor=blue            %color of external links
}

\begin{document}
\title{Congruence Kinematics in Conformal Gravity}

\author{Mohsen Fathi}
\email{m.fathi@shargh.tpnu.ac.ir; \,\,mohsen.fathi@gmail.com}

 \affiliation{Department of Physics, 
Payame Noor University (PNU),
P.O. Box 19395-3697 Tehran, Iran}

%%% Here starts the abstract
\begin{abstract}
In this paper we calculate the kinematical quantities of the
Raychaudhuri equations, to characterize a congruence of time-like
integral curves, according to the vacuum radial solution of Weyl
theory of gravity. Also the corresponding flows are plotted for
definite values of constants.
\bigskip

\noindent{\textit{keywords}}: Congruence kinematics, Raychaudhuri equations,
Weyl Gravity
\end{abstract}

\maketitle

\section{Introduction}
An important and notable feature of solutions to Einstein theory of gravity,
such as Schwarzchild spacetime or the cosmological
Friedmann-Lema\^{i}tre-Robertson-Walker solution, is that they all have apparent
singularities. The question was what would be their physical
implications? It was Raychaudhuri who tried to investigate these
solutions in both cosmological \cite{Raychaudhuri1} and
gravitational contexts \cite{Raychaudhuri1}. Later, in 1955, he
proposed his famous equations \cite{Raychaudhuri3} which are known
as Raychaudhuri equations.

These equations appeared to be of great importance in describing
gravitational focusing and spacetime singularities, which are essentially explained by the so-called Focusing
Theorems. Assuming Einstein's equations and using energy conditions, it is established that time-like and null geodesics if they are initially converging, they will focus until reaching zero size in a finite time. One of the most important notions about singularities, as pointed out by Landau and Lifshitz
\cite{Landau}, is that a singularity would always imply focusing
of geodesics, while focusing itself does not imply a singularity.
Therefore in their work, the concept of geodesic focusing
(although not explicitly stated the same) was worked out. However,
they did not introduce shear and rotation which are inseparable
ingredients of Raychaudhuri equations.

Ever since the equations were published, they have been discussed
and analyzed in numerous frameworks of general relativity, quantum
field theory, string theory and relativistic cosmology. One should
note that the concept of a singularity was first defined in
seminal works of Penrose and Hawking \cite{Penrose, Hawking1,
Hawking2}. Therefore it was only then that Raychaudhuri equations
received their deserved acclaim.

In addition to general relativity, the Raychaudhuri equations have
also been discussed in the context of alternative theories of
gravity, such as $f(R)$ theories \cite{Albareti}. In this paper as
well, we consider characteristics of time-like flows in an
alternative theory of gravity which as well has cosmological
implications, namely the Weyl theory of gravity. The paper is
organized as follows: in section 2 we briefly introduce the
Raychaudhuri equations and its kinematical parameters; in section
3 we mention the Weyl field equations and obtain the kinematical
evolution of time-like radial and rotational geodesic flows and
see how they behave by plotting the congruences of the integral
curves. Finally in section 4, we summarize the results.

\section{Raychaudhuri Equations}
As it was mentioned, the Raychaudhuri equations are evolution
equations for expansion, shear and rotation of a time-like
geodesic congruence of integral curves, which is indeed pure
geometrical and therefore, are independent of reference frames of
Einstein equations. Being parametrized in terms of the affine
parameter of the geodesics, $\tau$, these in
4-dimensional spacetimes are written as follows \cite{Wald}:
\begin{equation}\label{1}
    \frac{d\Theta}{d\tau}+\frac{1}{3}
    \Theta^2+\sigma^2-\omega^2=-R_{\mu\nu} v^\mu v^\nu,
\end{equation}
\begin{equation}\label{2}
    \frac{d\sigma_{\mu\nu}}{d\tau}=-\frac{2}{3} \Theta
    \sigma_{\mu\nu}-\sigma_{\mu\lambda}{\sigma^\lambda}_\nu-\omega_{\mu\lambda}{\omega^\lambda}_\nu+\frac{1}{3}h_{\mu\nu}\left(\sigma^2-\omega^2\right)
    $$$$+C_{\lambda\nu\mu\rho}v^\lambda v^\rho + \frac{1}{2}
    h_{\mu\lambda}h_{\nu\rho}R^{\lambda\rho}-\frac{1}{3}h_{\mu\nu}h_{\lambda\rho}R^{\lambda\rho},
\end{equation}
\begin{equation}\label{3}
    \frac{d\omega_{\mu\nu}}{d\tau}=-\frac{2}{3}\Theta\omega_{\mu\nu}-2{\sigma^\lambda}_{[\nu}\omega_{\mu]\lambda}.
\end{equation}
In Eqs. (\ref{1}) to (\ref{3}), $\Theta$, $\sigma_{\mu\nu}$ and
$\omega_{\mu\nu}$ are respectively the scalar expansion, the
symmetric trace-less shear tensor and the anti-symmetric rotation
tensor. Moreover $\sigma^2=\sigma_{\mu\nu}\sigma^{\mu\nu}$,
$\omega^2=\omega_{\mu\nu}\omega^{\mu\nu}$ and
$C_{\lambda\nu\mu\rho}$ is the Weyl conformal tensor. In Eq.~(\ref{1}), $v^\mu$ denotes the tangential vector field on the
geodesics and $h_{\mu\nu}$ in Eqs. (\ref{2}) and (\ref{3}) is the
projection tensor which for time-like curves and is defined by
\begin{equation}\label{4}
    h_{\mu\nu}=g_{\mu\nu}+v_\mu v_\nu.
\end{equation}
Generally speaking, the Raychaudhuri equations deal with the
kinematics of flows which are generated by vector fields. Such
flows are indeed congruences of integral curves which may or may
not be geodesics. Actually in the context of these equations, we
are interested in the evolution of the kinematical characteristics
of the so-called flows, not the origin of them. These
characteristics which are contained in the Raychaudhuri equations, may constitute one equation like
\cite{Kar2007}
\begin{equation}\label{5}
    \nabla_\nu v_\mu=\sigma_{\mu\nu}+\omega_{\mu\nu}+\frac{1}{3}
    h_{\mu\nu} \Theta,
\end{equation}
in which the trace-less symmetric part is defined as
\begin{equation}\label{6}
    \sigma_{\mu\nu}=
    \nabla_{(\nu}v_{\mu)}-\frac{1}{3}h_{\mu\nu}\Theta.
\end{equation}
Also the scalar trace is and the anti-symmetric part are
\begin{equation}\label{7}
    \Theta = \nabla_\mu v^\mu,
\end{equation}
\begin{equation}\label{8}
    \omega_{\mu\nu}=\nabla_{[\nu}v_{\mu]}.
\end{equation}
Geometrically, these quantities are related to a cross-sectional
area which encloses a definite number of integral curves and is
orthogonal to them. Moving along the flow lines, this area may
isotropically changes its size or being sheared or twisted,
however it still holds the same number of flow lines. There are
some analogies with elastic deformations which are discussed in Ref.~\cite{Poisson}. Also one can find explicit discussions on these
quantities in Refs.~\cite{Ellis, Ciufolini}.

Here we should note that the Raychaudhuri equations may be
essentially regarded as identities, which become equations when
they are for example used in spacetimes defined by Einstein field
equations.

Moreover, these equations are of first order and non-linear. Also
the expansion equation in Eq.~(\ref{1}), is the same as Riccati equation
in a mathematical point of view \cite{Tipler1, Tipler2}. The expansion is
indeed the rate of change of the cross-sectional area which is orthogonal
to the geodesic bundle.

In the next section, we will find the mentioned kinematical
characteristics, for curve bundles on a definite spacetime
background.

\section{Time-like Geodesic Congruences in Weyl Gravity}
The Weyl theory of gravity has had an interesting background and had received a great deal of attention from those who believe that the dark matter/dark energy scenarios could be well-treated by altering general theory of relativity. This became more elaborated after arguing that the extraordinary behavior of the galactic rotation curves could be extracted from Weyl gravity as a natural consequence of its vacuum solutions \cite{Mannheim,ja1,Kazanas}. There were subsequently more attempts to make relations between the theory's anticipations and observational evidences \cite{tekin1,fri,fri1} and as the main course, proposing gravitational alternatives to dark matter/dark energy \cite{Mannheim2005,Mannheim2006,Mannheim2007}. Some good information about this 4th order theory of gravity can be found in Ref. \cite{HWE}. Another important feature of the Weyl theory of gravity, is its conformal invariance which, as it is stated in the literature, could be considered as a tool of unification with the standard model by creating the desired mass during the symmetry breaking \cite{Pawlowski}.\\

The Weyl theory of gravity is a theory of 4th order with
respect to the metric. Weyl gravity is characterized by the Bach
action
\begin{equation}\label{9}
I_{B}=-\alpha\int{d^4x}\sqrt{-g}\,\,C^2,
\end{equation}
where $C^2=C_{\mu\nu\rho\lambda}C^{\mu\nu\rho\lambda}$ is the Weyl
invariant and $\alpha$ is a coupling constant. The action in Eq.~(\ref{9}) in principle, could be rewritten as
\begin{equation}\label{10}
I_{B}=-\alpha\int{\textmd{d}^4x}\sqrt{-g}\,\,\left(R^{\mu\nu\rho\lambda}R_{\mu\nu\rho\lambda}-2R^{\mu\nu}R_{\mu\nu}+\frac{1}{3}R^2\right)
\end{equation}
from which, using the total divergency of the Gauss-Bonnet term
$\sqrt{-g}\,\left(R^{\mu\nu\rho\lambda}R_{\mu\nu\rho\lambda}-4R^{\mu\nu}R_{\mu\nu}+R^2\right)$,
we have
\begin{equation}\label{11}
I_{B}=-\alpha\int{\textmd{d}^4x}\sqrt{-g}\,\,\left(R^{\mu\nu}R_{\mu\nu}-\frac{1}{3}R^2\right).
\end{equation}
Varying Eq.~(\ref{11}) with respect to $g_{\mu\nu}$, one obtains the
Bach tensor as \cite{Mannheim}
\begin{equation}\label{12}
W_{\mu\nu}=\nabla^\rho\nabla_\mu R_{\nu\rho}+\nabla^\rho\nabla_\nu
R_{\mu\rho}-\Box R_{\mu\nu}-g_{\mu\nu}\nabla_\rho\nabla_\lambda
R^{\rho\lambda}$$$$ -2R_{\rho\nu}
R^{\rho}_\mu+\frac{1}{2}g_{\mu\nu}R_{\rho\lambda}R^{\rho\lambda}-\frac{1}{3}\Big(2\nabla_\mu\nabla_\nu
R-2g_{\mu\nu}\Box R-2RR_{\mu\nu}+\frac{1}{2}g_{\mu\nu}R^2\Big).
\end{equation}
Accordingly, the Weyl field equations read as
\begin{equation}\label{13}
W_{\mu\nu}=\frac{1}{4\alpha} T_{\mu\nu},
\end{equation}
where $T_{\mu\nu}$ is the matter/energy tensor. Also the vacuum
field equations ($W_{\mu\nu}$=0) have been explicitly solved. The
solution constructs a spherically symmetric spacetime defined by
the line element \cite{Mannheim}
\begin{equation}\label{14}
ds^2=-B(r)dt^2+B^{-1}(r)dr^2+r^2d\Omega^2,
\end{equation}
in which
\begin{equation}\label{15}
B(r)=-\frac{\beta(2-3\beta\gamma)}{r}+(1-3\beta\gamma)+\gamma r-k
r^2.
\end{equation}
This solution has three important constants, $\beta, \gamma$ and
$k$, by which the Schwarzschild-de Sitter metric could be
regenerated. Also $\gamma$ and $k$ are respectively related to the
dark matter and dark energy constituents of the cosmic fluid. Now
let us inspect how time-like flows evolve in the spacetime
defined in Eq.~(\ref{14}). To do this, we separately consider radial
and rotational flows, and obtain the kinematical parameters in the
Raychaudhuri equations.

\subsection{Radial Flows}
\begin{figure}[t]
  \center{\includegraphics[width=6cm]{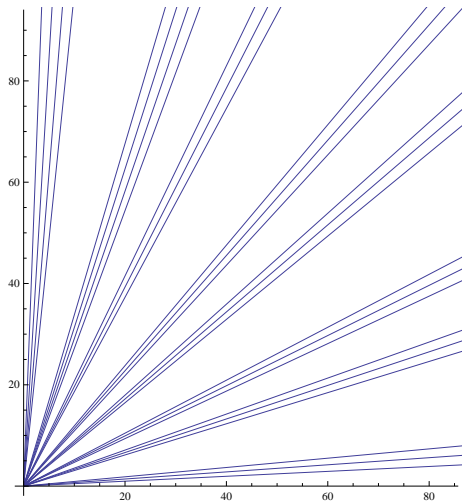}\\
  \caption{\label{fig:1}\small{The purely radial flows in a Weyl field with $\gamma=0$.}}}
\end{figure}

Some features of pure radial time-like flows has been discussed in Ref. \cite{Mohseni}. Firstly, for the spacetime coordinates on a parametric integral
curve
\begin{equation}\label{16}
    x^\mu=\left(t(\tau), r(\tau), \theta(\tau), \phi(\tau)\right),
\end{equation}
we define the velocity 4-vector field
\begin{equation}\label{17}
    v^\mu=\left(\dot t(\tau), \dot r(\tau), \dot\theta(\tau), \dot\phi(\tau)\right),
\end{equation}
which is supposed to be tangential to any integral curve in the
spacetime and for time-like congruences, one must have $v_\mu
v^\mu=-1$. Also $\dot x^\mu$ denotes $\frac{d}{d\tau}$. To have
purely radial flows on the equatorial plane, we take
$\theta=\frac{\pi}{2}$ and $\dot\phi=0$. Therefore for such flows
in the spacetime defined in Eqs.~(\ref{14}) and (\ref{15}), the
time-like condition is
\begin{equation}\label{18}
    g_{\mu\nu} v^\mu v^\nu=-1,
\end{equation}
and the geodesic equations
\begin{equation}\label{19}
    \dot v^\mu+{\Gamma^\mu}_{\nu\lambda} v^\nu v^\lambda=0,
\end{equation}
are respectively
\begin{equation}\label{20}
  \frac{r \dot r^2}{\beta  (3 \beta  \gamma -2)-k r^3+\gamma  r^2-3 \beta  \gamma
  r+r}+
  \frac{\dot t \left(2 \beta +k r^3-\gamma  \left(3 \beta ^2+r^2-3 \beta  r\right)-r\right)+r}{r}
  =0,
\end{equation}
\begin{equation}\label{21}
    \frac{\dot r \dot t \left(\beta  (3 \beta  \gamma -2)+2 k r^3-\gamma  r^2\right)}{r \left(\beta  (2-3 \beta  \gamma )+k r^3-\gamma  r^2
    +r (3 \beta  \gamma -1)\right)}+\ddot t=0,
\end{equation}
\begin{equation}\label{22}
    \frac{\left(\beta  (3 \beta  \gamma -2)+2 k r^3-\gamma  r^2\right) \left(\dot t^2 \left(\beta  (3 \beta  \gamma -2)-k r^3
    +\gamma  r^2-3 \beta  \gamma  r+r\right)^2-r^2 \dot r^2\right)}{2 r^3 \left(\beta  (2-3 \beta  \gamma )
    +k r^3-\gamma  r^2+r (3 \beta  \gamma -1)\right)}+\ddot r=0.
\end{equation}
Integrating Eq.~(\ref{21}) we obtain
\begin{equation}\label{23}
    \dot t=\frac{r}{-3 \beta ^2 \gamma +2 \beta +k r^3-\gamma  r^2+3 \beta  \gamma
    r-r},
\end{equation}
using which in Eq.~(\ref{20}), gives
\begin{equation}\label{24}
   \dot r= \pm\sqrt{\frac{-3 \beta ^2 \gamma +2 \beta +k r^3+3 \beta  \gamma  r-\gamma  r^2}{r}}.
\end{equation}
Taking the positive part, the tangential vector field in Eq.~(\ref{17})
for pure radial flows becomes
\begin{equation}\label{25}
   v^\mu=\left(\frac{r}{-3 \beta ^2 \gamma +2 \beta +k r^3-\gamma  r^2+3 \beta  \gamma  r-r}, \sqrt{\frac{-3 \beta ^2 \gamma +2 \beta +k r^3+3 \beta  \gamma  r-\gamma  r^2}{r}},0,0\right).
\end{equation}
The flows which are formed by this vector field, are expanding by
the following factor which is obtained by use of Eq.~(\ref{7}):
\begin{equation}\label{26}
    \Theta=\frac{6 k r^3-5 \gamma  r^2+3 \beta  (-3 \beta  \gamma +4 \gamma  r+2)}{2 r^{3/2} \sqrt{k r^3-\gamma  r^2+\beta  (-3 \beta  \gamma +3 \gamma
    r+2)}}.
\end{equation}
The vector field in Eq.~(\ref{25}) is orthogonal to the hypersurface of the
field crests. Therefore, it is of zero rotation \cite{Poisson}.
However, the shear tensor is non-zero. From Eqs.~(\ref{4}), (\ref{6})
and ({\ref{26}}), we have
\begin{subequations}\label{27}
\begin{align}
    \sigma_{tt}=\frac{\left(9 \beta ^2 \gamma +\gamma  r^2-6 \beta  (\gamma  r+1)\right) \sqrt{-3 \beta ^2 \gamma +k r^3-\gamma  r^2+\beta  (3 \gamma  r+2)}}{3 r^{5/2}},
\\
\sigma_{tr}=\sigma_{rt}=-\frac{9 \beta ^2 \gamma +\gamma  r^2-6
\beta  (\gamma  r+1)}{3 r \left(\beta  (2-3 \beta  \gamma )+k
r^3-\gamma  r^2+r (3 \beta  \gamma -1)\right)},
\\
\sigma_{rr}=\frac{\sqrt{r} \left(9 \beta ^2 \gamma +\gamma  r^2-6
\beta  (\gamma  r+1)\right)}{3 \sqrt{-3 \beta ^2 \gamma +k
r^3-\gamma  r^2+\beta  (3 \gamma  r+2)} \left(\beta  (3 \beta
\gamma -2)-k r^3+\gamma  r^2-3 \beta  \gamma  r+r\right)^2},
\\
\sigma_{\theta\theta}=\sigma_{\phi\phi}=-\frac{\sqrt{r} \left(9
\beta ^2 \gamma +\gamma  r^2-6 \beta  (\gamma  r+1)\right)}{6
\sqrt{-3 \beta ^2 \gamma +k r^3-\gamma  r^2+\beta  (3 \gamma
r+2)}}.
\end{align}
\end{subequations}
To obtain a pictorial viewpoint of how radial flows will behave in
a Weyl field, we should find an expression for $r(\tau)$. Using
Eq.~(\ref{23}), the geodesic equation in Eq.~(\ref{22}) gives
\begin{equation}\label{28}
    \ddot r-\frac{\left(\dot r^2-1\right) \left(\beta  (3 \beta  \gamma -2)+2 k r^3
    -\gamma  r^2\right)}{2 r \left(\beta  (2-3 \beta  \gamma )+k r^3-\gamma  r^2+r (3 \beta  \gamma
    -1)\right)}=0.
\end{equation}
Unfortunately, this is a non-linear second order equation. Even the
first order equation in Eq.~(\ref{24}) could not be explicitly solved.
Therefore, we consider a simpler case of $\gamma=0$ which is in
accordance to the de Sitter solution. In this case, Eq.~(\ref{24})
results in
\begin{equation}\label{29}
    r(\tau)=\pm\frac{\textrm{e}^{-\sqrt{k} (\text{c}+\tau )} \left(e^{3 \sqrt{k} (\text{c}+\tau )}-2 \beta  k\right)^{2/3}}{2^{2/3}
    k^{2/3}},
\end{equation}
with
$$c=\frac{1}{3}\frac{\ln \left({2}{\sqrt{k^3 {r_0}^3 \left(2 \beta +k {r_0}^3\right)}+k^2 {r_0}^3+\beta  k}\right)}{\sqrt{k}},$$
where $r_0$ is the point of closest approach. The radial flows are
shown in Fig.~\ref{fig:1}. One can note that how the expansion will make
cross-sectional area to change its size.

\subsubsection{Focusing}

If spacetime singularities are concerned, Eq.~(\ref{1}) for
the expansion receives the most central attention. As it was shown
above (and also in Fig.~\ref{fig:1}), the expansion changes the
cross-sectional area along the geodesic bundle. It is noted that,
if the expansion approaches the negative infinity, the congruences
will converge, whereas they diverge if the expansion goes to
positive infinity. The convergence, is what we regard as the focusing
of the geodesic bundles toward the singularities. However,
to clarify whether any convergence occurs for a peculiar flow, one
should examine the expansion equation. It has been put in the
literature that, convergence occurs if
\begin{equation}\label{29-1}
    R_{\mu\nu} v^\mu v^\nu + \sigma^2-\omega^2 \geq0,
\end{equation}
or equivalently,
\begin{equation}\label{29-2}
    \frac{d\Theta}{d\tau}+\frac{1}{3}\Theta^2\leq0.
\end{equation}
Therefore one can observe that shear acts in favor of convergence
while rotation opposes it. However, for zero rotations the
condition in Eq.~(\ref{29-1}) reduces to \cite{Kar2007}
\begin{equation}\label{29-3}
    R_{\mu\nu} v^\mu v^\nu \geq0.
\end{equation}
Now, for the zero-rotational flow defined by the vector field in Eq.~(\ref{25}), the condition in Eq.~(\ref{29-3}) gives
\begin{equation}\label{29-4}
   \frac{\gamma }{r}-3 k \geq0.
\end{equation}
This provides us a maximum for $r$ according to which, one must
have $r\leq\frac{\gamma}{3k}$. We should note that the metric
potential in Eq.~(\ref{15}), includes a Newtonian $\frac{1}{r}$ term
which is dominant at small distances. Increasing $r$, it would be
the term $\gamma r$ as the dominant one. Such distances are about
galactic scales. For a typical galaxy of of radius $r\sim10
\textrm{kpc}$, $\gamma$ is of order $10^{-26} \textrm{m}^{-1}$
and at cosmological distances, the coefficient $k$ in the term $k
r^2$ is of the greatest importance which for a universe of constant
curvature, may be regarded as the cosmological parameter
of order $10^{-43} \textrm{m}^{-2}$ \cite{Varieschi2014,Mureika2017}. Therefore if galactic scales are of interest, Eq.~(\ref{29-4}) implies that for $r\lesssim3.3\times 10^{16}\textrm{m}$ we may expect the convergence of the flow.

\subsection{Rotational Flows}
\begin{figure}[t]
  \center{\includegraphics[width=6cm]{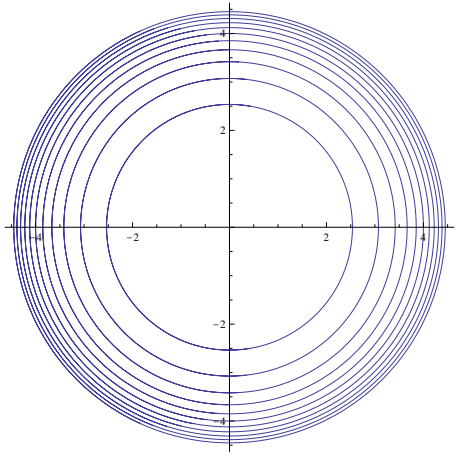}\\
  \caption{\label{fig:2}\small{The purely rotational flows in a Weyl field.}}}
\end{figure}

Pure rotational flows in the equatorial plane, could be obtained by
letting $\theta=\frac{\pi}{2}$ and $r=\textrm{const}.$, according
to the vector field in Eq.~(\ref{17}). Hence, the time-like
condition in Eq.~(\ref{18}) and the geodesic Eqs.~(\ref{19}) are
respectively
\begin{equation}\label{30}
    \dot t^2 \left(\beta  (2-3 \beta  \gamma )+k r^3-\gamma  r^2+r (3 \beta  \gamma -1)\right)+r^3 \dot \phi^2+r=0,
\end{equation}
\begin{equation}\label{31}
    \ddot t=0,
\end{equation}
\begin{equation}\label{32}
   \dot t^2 \left(\beta  (3 \beta  \gamma -2)+2 k r^3-\gamma  r^2\right)+2 r^3 \dot \phi^2=0,
\end{equation}
\begin{equation}\label{33}
    \ddot \phi=0.
\end{equation}
Equation (\ref{31}) implies $\dot t=1$. Applying this to
Eq.~(\ref{30}) gives
\begin{equation}\label{34}
    \dot \phi=\pm\frac{\sqrt{3 \beta ^2 \gamma -2 \beta -k r^3+\gamma  r^2-3 \beta  \gamma
    r}}{r^{3/2}}.
\end{equation}
Therefore, the tangential vector field can be written as
\begin{equation}\label{35}
    v^\mu=\left(1,0,0,\frac{\sqrt{3 \beta ^2 \gamma -2 \beta -k r^3+\gamma  r^2-3 \beta  \gamma
    r}}{r^{3/2}}\right).
\end{equation}
Together with Eq.~(\ref{7}), one can see that $\Theta=0$, implying
that the flows which are formed by Eq.~({\ref{35}}), are free of
expansion. Therefore we can not expect any focusing for this
rotational flow, since the cross-sectional area does not change.
Also the non-zero components of the shear tensor would be
\begin{equation}\label{36}
    \sigma_{r\phi}=\sigma_{\phi r}=\frac{-9 \beta ^2 \gamma -\gamma  r^2
    +6 \beta  (\gamma  r+1)}{4 \sqrt{r} \sqrt{3 \beta ^2 \gamma -k r^3+\gamma  r^2-\beta  (3 \gamma  r+2)}}.
\end{equation}
Moreover, using Eq.~(\ref{8}), one can obtain the rotation of the flow
which is characterized by the following non-zero components of the
anti-symmetric part of the Raychaudhuri kinematical parameters:
\begin{subequations}\label{37}
\begin{align}
    \omega_{tr}=-\omega_{rt}=k r-\frac{-3 \beta ^2 \gamma +2 \beta +\gamma  r^2}{2 r^2},
    \\
    \omega_{r\phi}=-\omega_{\phi r}=\frac{-3 \beta ^2 \gamma
    +4 k r^3-3 \gamma  r^2+\beta  (6 \gamma  r+2)}{4 \sqrt{r} \sqrt{3 \beta ^2 \gamma -k r^3+\gamma  r^2-\beta  (3 \gamma  r+2)}}.
 \end{align}
\end{subequations}
The constant radial distance can be obtained from Eqs.~(\ref{32}) and
(\ref{34}). We have
\begin{equation}\label{38}
    r=\frac{3 \beta  \sqrt{\gamma }\pm\sqrt{6} \sqrt{\beta }}{\sqrt{\gamma}}.
\end{equation}
According to this, the pure rotational flow can be obtained which
has been depicted in Fig.~\ref{fig:2}. One can note that the
cross-sectional area will suffer a twist while holding the same
number of integral curves. However for each curve, the radial
distance is a constant and the shear gradually will make the
congruence bundle to become compressed.

\subsection{Radial-Directional Flow}

Now let us consider a geodesic flow, for which both radial and
angular components of the tangent vector field of the congruence
are supposed to be affine variables. We are still on the
equatorial plane, so the time-like condition (\ref{18}) and the
geodesic Eqs.~(\ref{19}) read as
\begin{equation}\label{39}
    \dot t^2 \left(\frac{k r^3-\gamma  r^2+\beta  (-3 \beta  \gamma +3 \gamma  r+2)}{r}-1\right)
    +\frac{r \dot r^2}{\beta  (3 \beta  \gamma -2)-k r^3+\gamma  r^2-3 \beta  \gamma  r+r}+r^2
    \dot\phi^2=-1,
\end{equation}
\begin{multline}\label{40'}
    \ddot r +\frac{1}{2 r^3 \left(\beta  (2-3 \beta  \gamma )+k r^3-\gamma  r^2+r (3 \beta  \gamma
    -1)\right)}
   \\
    \times\Big[\left(\beta  (3 \beta  \gamma -2)-k r^3+\gamma  r^2-3 \beta  \gamma  r+r\right)^2 \left(\dot t^2 \left(\beta  (3 \beta  \gamma -2)+2 k r^3-\gamma  r^2\right)
    +2 r^3 \dot\phi^2\right)\\
   +r^2 \dot r^2 \left(\beta  (2-3 \beta  \gamma )-2 k r^3+\gamma
    r^2\right)\Big]=0,
   \end{multline}
   \begin{equation}\label{40}
    \frac{2 \dot r \dot\phi}{r}+\ddot\phi=0.
    \end{equation}
The temporal part of the geodesic equations and consequently $\dot t$, are the
same as those Eqs.~(\ref{21}) and (\ref{23}). Also direct integration of
$\phi$ equation in (\ref{40}) gives
\begin{equation}\label{41}
    \dot \phi =\left(\frac{1}{r}\right)^2.
\end{equation}
Using Eqs.~(\ref{23}) and (\ref{41}) in Eq.~(\ref{39}) to obtain $\dot r$,
the tangential vector field for the geodesic congruence becomes
\begin{multline}\label{42}
    v^\mu=\left(\frac{r}{-3 \beta ^2 \gamma +2 \beta +k r^3-\gamma  r^2+3 \beta  \gamma  r-r},\right.\\
    \left. {\sqrt{\frac{1}{r}\left[\beta  (3 \beta  \gamma -2)-k r^3+\gamma  r^2-3 \beta  \gamma  r+r\right]} \sqrt{\frac{r}{\beta  (3 \beta  \gamma -2)-k r^3+\gamma  r^2-3 \beta  \gamma
    r+r}-\frac{1}{r^2}-1}}{{}}
    , 0, \frac{1}{r^2}\right).
\end{multline}
According to Eq.~(\ref{42}) one can obtain the kinematical
characteristics of a Radial-Directional flow in a Weyl field. The
expansion is
\begin{equation}\label{43}
\Theta=\frac{1}{2 r^{7/2}A}\Big[\beta  (2-3 \beta  \gamma )+6 k
r^5+4 r^3 (3 \beta \gamma +k)-5 \gamma  r^4-3 r^2 (\beta  (3 \beta
\gamma -2)+\gamma )+r (6 \beta  \gamma -2)\Big],
\end{equation}
where
$$A=\sqrt{\beta  (3 \beta  \gamma -2)-k r^3+\gamma  r^2-3 \beta  \gamma  r+r} \sqrt{\frac{r}{\beta  (3 \beta  \gamma -2)-k r^3
+\gamma  r^2-3 \beta  \gamma  r+r}-\frac{1}{r^2}-1}.$$
Moreover, the non-zero components of the shear tensor become
\begin{subequations}\label{44}
\begin{align}
\sigma_{tt}=\frac{1}{6 r^{9/2} A}\times
     \Big[-4 \beta ^2 (2-3 \beta  \gamma )^2+2 k^2 r^6+2 k r^3 \left(\beta  (3 \beta  \gamma -2)+\gamma  r^4
    -6 \beta  \gamma  r^3+r^2 \left(9 \beta ^2 \gamma -6 \beta -\gamma \right)-2 r\right)
    \nonumber\\-2 \gamma ^2 r^6+18 \beta  \gamma ^2 r^5
    +4 \beta  \gamma  r^4 (4-15 \beta  \gamma )+\gamma  r^3 \left(90 \beta ^3 \gamma -60 \beta ^2+6 \beta  \gamma +1\right)
    \nonumber\\-2 \beta  r^2 \left(27 \beta ^3 \gamma ^2-36 \beta ^2 \gamma +3 \beta  \left(5 \gamma ^2+4\right)
    -7 \gamma \right)+\beta  r \left(54 \beta ^2 \gamma ^2-51 \beta  \gamma
    +10\right)\Big],
    \\\nonumber\\
\sigma_{tr}=\sigma_{rt}=\frac{\beta  (2-3 \beta  \gamma )+4 r^3 (3 \beta  \gamma +k)-2 \gamma  r^4
    -3 r^2 \left(6 \beta ^2 \gamma -4 \beta +\gamma \right)+r (6 \beta  \gamma -2)}{6 r^3 \left(\beta  (2-3 \beta  \gamma )
    +k r^3-\gamma  r^2+r (3 \beta  \gamma -1)\right)},    
    \\\nonumber\\
    \sigma_{t\phi}=\sigma_{\phi t}=-\frac{1}{6 r^{7/2}A}\Big[\beta  (2-3 \beta  \gamma )+6 k r^5
    +4 r^3 (3 \beta  \gamma +k)-5 \gamma  r^4-3 r^2 (\beta  (3 \beta  \gamma -2)+\gamma )+r (6 \beta  \gamma -2)\Big],
    \\\nonumber\\
    \sigma_{rr}=\frac{1}{6 r^{9/2}\left(\beta  (3 \beta  \gamma -2)-k r^3+\gamma  r^2-3 \beta  \gamma  r+r\right)^{2} A}\times\Big[-\beta ^2 (2-3 \beta  \gamma )^2+r^6 \left(\gamma  (\gamma -12 \beta )-4 k^2
    +2 k (3 \beta  \gamma -5)\right)-\gamma (k-2) r^7\nonumber\\+r^5 \left(-9 \beta ^2 \gamma  (k-2)+3 \beta  \left(-3 \gamma ^2+2 k
    -4\right)+\gamma  (7 k+10)\right)+r^4 \left(30 \beta ^2 \gamma ^2-38 \beta  \gamma -3 \gamma ^2+k (6-18 \beta  \gamma )
    +8\right)\nonumber\\-5 r^3 \left(9 \beta ^3 \gamma ^2+\gamma -3 \beta ^2 \gamma  (k+4)+\beta  \left(-3 \gamma ^2+2 k+4\right)\right)
    \nonumber\\+r^2 \left(27 \beta ^4 \gamma ^2-36 \beta ^3 \gamma -6 \beta ^2 \left(5 \gamma ^2-2\right)
    +20 \beta  \gamma -2\right)+3 \beta  r \left(9 \beta ^2 \gamma ^2-9 \beta  \gamma
    +2\right)\Big],
    \\\nonumber\\
    \sigma_{r\phi}=\sigma_{\phi r}=\frac{\beta  (2-3 \beta  \gamma )+r^3 (-6 \beta  \gamma +4 k+6)+\gamma  r^4+r^2 \left(9 \beta ^2 \gamma -6 \beta
    -3 \gamma \right)+r (6 \beta  \gamma -2)}{6 r^3 \left(\beta  (2-3 \beta  \gamma )
    +k r^3-\gamma  r^2+r (3 \beta  \gamma -1)\right)},
    \\\nonumber\\
     \sigma_{\theta\theta}=\frac{1}{6 r^{3/2}A}\Big[5 \beta  (2-3 \beta  \gamma )+2 r^3 (3 \beta  \gamma +k)
    -\gamma  r^4-3 r^2 \left(3 \beta ^2 \gamma -2 \beta +\gamma \right)+4 r (3 \beta  \gamma
    -1)\Big],
    \\\nonumber\\
   \sigma_{\phi\phi}=\frac{1}{6 r^{7/2} A}\times\Big[
    \beta  (3 \beta  \gamma -2)-3 k r^7-r^5 (3 \beta  \gamma +7 k)-r^3 (9 \beta  \gamma +4 k+1)\nonumber\\+3 r^2 \left(\beta  (2-3 \beta  \gamma )
    +k r^5+r^3 (3 \beta  \gamma +k)-\gamma  r^4-r^2 \left(3 \beta ^2 \gamma -2 \beta +\gamma \right)
   +r (3 \beta  \gamma -1)\right)
    \nonumber\\+2 \gamma  r^6+5 \gamma  r^4+r^2 \left(3 \beta ^2 \gamma -2 \beta +3 \gamma \right)+r (2-6 \beta  \gamma )\Big].  
\end{align}
\end{subequations}
Surprisingly, the rotation tensor vanishes;
$\omega_{\mu\nu}=\textbf{0}$. Therefore, if one is interested in
focusing, it is sufficient to examine Eq.~(\ref{29-3}). For
Eq.~(\ref{42}) this gives the condition
\begin{equation}\label{45}
  {3 \beta  \gamma -\left(r^2+1\right) r (3 k r-\gamma )+3 k r^2-2 \gamma
   r}\geq0.
\end{equation}
Once again, dealing with a typical galaxy, the dimension-less term
$\beta\gamma$ would be of order $10^{-12}$. Counting on our previous values for $\gamma$ and $k$, then the
only non-zero solution for $r$ is
\begin{equation}\label{46}
    r= 3.33333\times 10^{16} \textrm{m}.
\end{equation}
Note also that, this is in agreement with the convergence
condition for the purely radial flows which was discussed above.

\subsubsection{Capture Zone}

\begin{figure}[t]
  \center{\includegraphics[width=8cm]{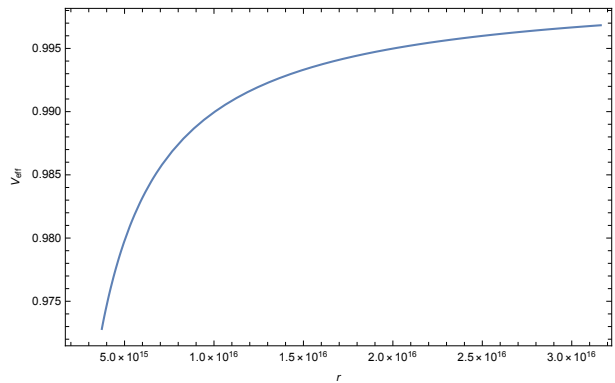}\\
  \caption{\label{fig:3}\small{The effective potential for a test particle moving on a radial-directional time-like geodesic in a Weyl field. Plotting has been done for $\gamma=10^{-26} \textrm{m}^{-1}$, $\beta=10^{14} \textrm{m}$ and $k=10^{-43} \textrm{m}^{-2}$.}}}
\end{figure}

One might even obtain the region in which there would be a
possible capturing of the flow, by use of an effective potential.
To proceed with this method, let us use the time-like condition in Eq.~(\ref{39}), exploiting the total energy definition $E=g_{00}\,
\dot t$ and also Eq.~(\ref{41}). Rearrangement yields
\begin{equation}\label{47}
    \dot r^2=E^2 - {V_{\textrm{eff}}(r)}^2,
\end{equation}
where
\begin{equation}\label{48}
    V_{\textrm{eff}}(r)=\sqrt{\frac{\left(r^2+1\right) \left(\beta  (3 \beta  \gamma -2)-k r^3+\gamma  r^2-3 \beta  \gamma
    r+r\right)}{r^3}}.
\end{equation}
This potential has been plotted in Fig.~\ref{fig:3}. However in order to
capture a time-like flow, it is pretty useful to take care about
the potential maximums. Such maximums may represent unstable
circular orbits. Hence if the energy of a particle is supposed to
be the same as the potential maximum, it will be inevitably
captured and the corresponding time-like geodesics will ultimately
terminated where the potential originates from. The potential
in Eq.~(\ref{48}) has a maximum around $r \approx 3\times 10^{16} \textrm{m}$,
where the derivatives of $V_{\textrm{eff}}(r)$ vanish. So for this
maximum and higher energies, one can expect geodesic focusing.

\section{Summary}
In this paper we dealt with the characteristics of time-like geodesic
flows in a Weyl field and we supposed that Such field is formed in
vacuum spacetime, obtained from vacuum Weyl field equations. We
distinctly considered radial and rotational flows and obtained
their corresponding expansion, shear and vorticity (i.e. the
rotation tensor). We noted that for radial flows, only expansion and shear do contribute in the characterization and
as it is obvious in Fig.~\ref{fig:1}, the expansion is isotropic, however
the shear makes the formation of the curve bundle an-isotropic.
Moreover, for the peculiar rotational flows we considered, we
found that although it is not usual, however, the expansion
vanishes. Therefore we are left with a simple shear tensor and of
course a non-vanishing vorticity. This may provide us a circular
flow, which is gradually getting compact. For further
considerations, one may concern with null-like geodesics in Weyl
fields. Also it is possible to take both rotation and radial
motions in same flows. In cosmological contexts, this may help
us to discover how real cosmic flows will behave in Weyl conformal
gravity.

\end{document}